\documentclass[12pt]{article}
\usepackage{geometry}                
\usepackage{graphicx}
\usepackage{amssymb}
\usepackage{amsmath}
\usepackage{setspace}
\usepackage{authblk}
\title{Selection biases the prevalence and type of epistasis along
adaptive trajectories}
\author[1, 2, *]{Jeremy A. Draghi}
\author[1]{Joshua Plotkin}

\affil[1]{Department of Biology, University of Pennsylvania, Philadelphia, PA, USA}
\affil[2]{Present Address: Department of Zoology, University of British Columbia, Vancouver, BC, Canada}
\affil[*]{Corresponding author}

\begin{document}
\date{}
\maketitle
\clearpage
\section{Abstract}

The contribution to an organism's phenotype from one genetic locus may depend upon
the status of other loci. Such epistatic interactions among loci are now
recognized as fundamental to shaping the process of adaptation in evolving
populations. Although little is known about the structure of epistasis in most
organisms, recent experiments with bacterial populations have concluded that
antagonistic interactions abound and tend to de-accelerate the pace of adaptation
over time. Here, we use a broad class of mathematical fitness landscapes to
examine how natural selection biases the mutations that substitute during
evolution based on their epistatic interactions. We find that, even when beneficial mutations
are rare, these biases are strong and change substantially throughout the course
of adaptation. In particular, epistasis is less prevalent than the neutral expectation 
early in adaptation and much more prevalent later, with a concomitant
shift from predominantly antagonistic interactions early in adaptation to
synergistic and sign epistasis later in adaptation. We observe the same patterns
when re-analyzing data from a recent microbial evolution experiment.  Since these
biases depend on the population size and other parameters, they must be quantified
before we can hope to use experimental data to infer an organism's underlying
fitness landscape or to understand the role of epistasis in shaping its
adaptation.  In particular, we show that when the order of substitutions is not
known to an experimentalist, then standard methods of analysis may suggest that
epistasis retards adaptation when in fact it accelerates it.

\section{Author Summary}

A major goal in evolutionary biology is to relate the adaptation of an organismÕs traits to changes in their genome. When researchers identify substitutionsÑpopulation-wide changes in DNA sequenceÑthey often find that the effects of those changes on fitness interact epistatically, or non-linearly.  While epistasis is often assumed to be relevant only for very large and diverse populations, here we present computational results that demonstrate that epistasis shapes the pattern of adaptive change in all populations. Our results allow us to make predictions about what kind of epistasis is most likely to occur, with distinct predictions for patterns in the early or later phases of adaptation. We also apply these results to recent evolution experiment with microbes and show that epistasis may appear to slow adaptation, when in fact it may be facilitating evolution by natural selection. Our results will provide a new basis for comparison for all experimentalists who investigate the fitness effects of the molecular changes underlying adaptation. 

\clearpage

\section{Introduction}

Two sites in a genome interact epistatically when the contribution to a trait at
one site depends on the state of the other site. While epistasis has been a
significant theme in topics such as the evolution of sex and robustness to
mutation, its role in the dynamics of evolving populations has only begun to be
explored. Recent experimental evolution studies of microbes \cite{Blount:2008zi,
Chou:2011zr, Khan:2011ly, Woods:2011hc} and biomolecules \cite{Reetz:2008vn,
Bloom:2009rc, Hayden:2011ys, Salverda:2011bh} have revealed that epistasis is
widespread and consequential for adaptation. These studies, combined with
experiments that reconstruct ancestral genotypes
\cite{Weinreich:2006kx, Bridgham:2009fk, Lozovsky:2009oq, Bloom:2010fv,
Lunzer:2010uq, Novais:2010dq, Martinez:2011kl} or examine numerous combinations of
adaptive mutations \cite{Remold:2004ij, Trindade:2009cr, Kvitek:2011tg,
Rokyta:2011pi}, have amply demonstrated that molecular evolution
cannot be explained or predicted without understanding how gene interactions shape
adaptive possibilities. 
 
Evolution experiments are increasingly used to address two questions that are
conceptually distinct, but empirically entangled: of the mutations that could play
a role in adaptation, how do their interactions shape evolution, and how do
evolutionary processes determine what kinds of interactions occur among
the sites that substitute? The first question focuses on the properties of the
genotype-phenotype-fitness relationship, also called the adaptive landscape, while
the second question asks how the mutations chosen by evolution reflect that
underlying landscape. If selection and other evolutionary forces are blind to
epistasis -- that is, if interactions among sites do not influence the likelihood
that they will substitute -- then this second question is irrelevant, and the
genetic changes we see in evolution experiments perfectly mirror the
epistatic properties of the underlying adaptive landscapes. If, however, evolution
biases the fixation of groups of mutations with specific patterns of interactions,
then evolution experiments present a complex problem: if epistasis shapes
evolution, and evolution distorts the appearance of epistasis, then how can we use
evolution experiments to infer the underlying fitness landscape? This ambiguity
complicates even qualitative inferences such as whether epistasis among genes
can be said to have slowed or hastened adaptation. To resolve this ambiguity
researchers must first understand how evolution biases the combinations of sites
that substitute in an adapting population. Only then can researchers hope to
correct for these biases, which will depend upon the size, mutation rate, and
other characteristics of the population, in order to infer the underlying fitness
landscape from experimental data.

Many theoretical studies of epistasis and patterns of asexual adaptation have
focused on questions of the existence and accessibility of multiple fitness peaks
\cite{Kauffman:1993fk, WHITLOCK:1995ly, Weinreich:2005fk, Cowperthwaite:2006fv,
Weissman:2009dq, Carneiro:2010zr, Dawid:2010qf, Franke:2011ve, Ostman:2012uq}.
While such work has clarified the broad-scale picture of how epistasis shapes
adaptation, its usefulness in predicting microevolutionary dynamics is limited. In
contrast, our interest here is how experiments on adapting populations can be used
to infer the properties of an organism's underlying fitness landscape and how
epistasis shapes those experimental outcomes. 

We use a computational model to clarify the evolutionary effects of two
contradictory roles of epistasis: epistatic interactions can undermine the benefits of
previously adaptive genetic substitutions, but they can also produce new paths to
higher fitness \cite{Draghi:2011ve, Wagner:2011tg}. The first of these effects
would tend to retard adaptation, and the latter effect would accelerate it. Our results
show that natural selection biases the prevalence and type of epistatic
interactions among the mutations that substitute, even when mutations are too rare
to interact directly as coexisting polymorphisms.  Here we work to quantify
precisely how
selection biases the epistasis among mutations that substitute
in an adapting population and to understand why these biases arise.

\section{Results}

\subsection{Prevalence of epistasis along an adaptive walk}

To understand how selection shapes epistasis among the mutations that
substitute, we first simulated adaptive
walks in which beneficial mutations substitute sequentially. \O stman et al.
suggest that epistatic interactions should not influence substitution patterns
when rare beneficial mutations fix independently \cite{Ostman:2012uq}. Our results initially 
appear to confirm
this expectation: Figure 1(a) shows that sites that fix sequentially are almost as
likely to interact epistatically as are pairs of randomly chosen sites. However,
this concordance disappears when epistasis is examined along the sequence of steps
comprising an adaptive walk: Fig. 1(b) shows that epistasis is in fact suppressed
early in adaptation, and enriched among later steps, compared to a random
(neutral) walk. Thus, selection biases the amount of epistasis among the
mutations that fix along an adaptive walk, and it does so in a complex manner. These opposing
effects produce an apparent agreement with the random expectation for the overall
prevalence of epistasis when observations are coarsely averaged across entire
walks, but in fact these results demonstrate that epistasis can shape patterns of
substitutions even when mutations fix independently, one after another.

Why does selection suppress epistasis early in walks and promote it later? To
address this, we studied how mutations at sites along an adaptive walk influence
the fitness effects of the sites with which they interact (upstream or
downstream). In particular, in the simple case of $K = 1$, Fig.~2 shows the
distribution of fitness effects of mutations at those sites that do and do not interact
with the site that has just substituted along a
walk. For a site $i$ that changes early in the walk, mutations at its
interacting sites are less likely to be beneficial. In other words, adaptive
substitutions early in the walk partly undermine the benefits that would be
conferred by mutations at their partner sites. Therefore, after an early
substitution at one site, its epistatic partners are less likely to substitute
than they would have otherwise  -- and so the early steps in an adaptive walks
exhibit a deficit of epistasis compared to the neutral expectation.

This bias against epistasis early in a walk is caused by the dependency of
selected substitutions on the backgrounds in which they were selectively favored.
When a site forms part of the relevant genetic background for a beneficial
substitution, it is statistically likely that changes at such a site would
partially undermine the beneficial effect of this fixed adaptive substitution.
Supplemental Figure 6(a) demonstrates this regression to the mean effect:
selective coefficients of mutations at a site $j$ are suppressed when $j$
interacts with a site $i$ that has just fixed an adaptive mutation, and this
suppression is greater when the beneficial effect of the substitution at $i$ is
larger. 

\subsection{Form of epistasis along an adaptive walk}
Aside from biasing the amount of epistasis along a walk, selection also biases the
type of epistasis between successive substitutions. We find that the predominant
sign of epistasis, as well as its prevalence, depends on the position of the
substitutions along an adaptive walk. Fig.~3 shows that early substitutions tend
to show antagonism with one another, while later substitutions typically exhibit
sign epistasis, defined as pairs of mutations where at least one member has a
beneficial fitness effect on one background and a deleterious effect on the other.
Synergy between beneficial mutations is present but less common than antagonism at
early steps and less common than sign epistasis at later steps. The shift from
antagonistic towards synergistic/sign epistasis helps to explain why epistasis is
suppressed early in an adaptive walk and augmented later in the walk. 

Selection also biases the directionality of interactions between successive
substitutions along an adaptive walk (see Methods). Fig.~4a shows that
interactions with $i$ upstream of $j$ are more frequent than the converse along
the entire adaptive walk. This difference is explained by disparate effect on the
evolvability of site $j$: $j$ is more likely to become evolvable (able to
substitute beneficially) if it is influenced by the preceding substitution at $i$,
than if it influences $i$ (see Supplemental Text).  This result is confirmed by
the regressions in Supplemental Figure 6(b \& c), and it explains why such
interactions are more prevalent than the converse (Figure 4a). When $K > 1$,
another type of interaction is possible: $i$ and $j$ might not influence each
other directly, but both might influence a third site. Epistasis of this type is
expected to be very common when $K$ is a substantial fraction of $N$, and results
for $K = 5$ show that the prevalence of this type of interaction also changes
along adaptive walks (Fig.~4b). Thus, evolution biases the types and directions of interactions
among substitutions along an adaptive walk.

Choosing the $K$ sites which influence each of the $N$ loci defines a network of
directed interactions. While the number of sites that influence a locus (its
in-degree) is fixed at $K$, the number of sites that a locus influences (its
out-degree) is variable and approximately Poisson distributed. Evolutionary
preferences for certain types of epistasis, such as the bias in favor of
substitutions at sites that depend on previous substitutions, might then lead to
systematic variation in the out-degree of sites which substitute adaptively.
Supplemental Figure 7 confirms this hypothesis: the out-degree of substituted
sites is initially slightly higher than expected, then declines with substitution
number. This evidence implies that those sites with more epistatic connections
are less likely to change during adaptation, because mutations at such sites have
a greater chance of undermining previously selected beneficial changes. 

\subsection{Robustness of results and comparison to data}

The NK model has several features that could amplify the biases in epistasis
introduced by natural selection.  In order to assess whether these model
assumptions might lead to spurious results, we explored how the patterns shown in
Figure 4 change as the number of alleles per site and the starting fitnesses were
varied. Supplemental Figure 8 shows that deviations from the expected prevalence
of epistasis are qualitatively similar, and quantitatively greater, when the
number of alleles per site, $A$, is increased.  Supplemental Figure 9 confirms
that the basic pattern of our results is also robust to changes in the fitness of
the starting genotype.

While we have focused on evolutionary dynamics in the simplified,
strong-selection-weak-mutation regime, we can use individual-based simulations to
explore epistasis in polymorphic populations with larger values of the
population-scaled mutation rate, $\theta$. Supplemental Figures 10 and 11
show very similar patterns for the prevalence of epistasis among adaptive
substitutions, even when $\theta$ is much greater than 1. Some differences from
Fig.~4 are apparent in the high-$\theta$ case during the period of stabilizing
selection that follows adaptation. However, these simulations at high $\theta$
confirm that the major patterns found in adaptive walks at low $\theta$: there is
a deficit of epistasis early in evolution, and a surplus of epistasis later in
adaptation, with a predominance of interactions in which the effect of each
substitution depends on the preceding substitution along the line of descent.

We also considered a very different set of fitness landscapes -- computationally
predicted RNA folding -- to assess the generality of our principal findings. RNA
sequences do not have static epistatic interactions between sites; instead, interactions emerge from the folding topology and change with genotype.
However, we can still measure the average frequency of epistasis between
substitutions on an evolutionary line of descent; such data show that the
prevalence of epistasis does vary systematically along a series of substitutions,
although the trend is toward decreased epistasis (Fig.~7(a)). The type of epistasis
can also be quantified, although the high frequency of conditionally neutral
mutations in RNA necessitates a new category of neutral epistasis. Fig.~7(b)
shows that, as in the NK model, early antagonism gives way to a high prevalence of
sign (and neutral) epistasis later in adaptation. While these results differ in
some features from those obtained in the NK model, they further illustrate that
evolution at $\theta < 1$ can indeed bias the epistatic properties of fixed
mutations, and it does so differentially at different stages of adaptation. 

Finally, we re-examined data from the microbial evolution experiment of Khan et
al.  \cite{Khan:2011ly}, in which the actual order of substitutions that occurred
is known.  As shown in Supplemental Figure 15, we used their fitness measurements
to calculate epistasis along the path of adaptive change. Two methods of
calculating expected fitness, which differ in their choice of reference genotype,
both yield the same qualitative result: epistasis is initially negative, then
becomes positive during the later stages of adaptation.  While this pattern
represents only a single instance of an empirical evolutionary trajectory, its
similarity to the patterns expected under our analysis of a broad class of
mathematical fitness landscapes (Fig.~3) is striking. This re-analysis suggests
that epistasis may in fact be accelerating late adaptation in these experimental
populations, in contrast to the original interpretation of the data
\cite{Khan:2011ly, Kryazhimskiy:2011}.

\subsection{Implications for inferences from experimental data}

Two recent studies on experimental populations of bacteria have inferred that
antagonistic epistasis among beneficial mutations is common and ultimately
explains a trend of diminishing fitness gains over time \cite{Chou:2011zr,
Khan:2011ly}. These studies relied in part on regression analyses of the fitness
effects of observed substitutions in the presence and absence of the other
beneficial substitutions observed in the experiment. Both studies found a trend
towards smaller beneficial effects when substitutions were assayed in backgrounds
of higher fitness and so concluded that antagonistic epistasis decelerates
adaptation. However, we demonstrate below that in the NK model, such regressions
are not a reliable indicator of the effect of epistasis on the speed or extent of
adaptation. Our results suggest that a common statistical artifact -- regression
to the mean -- confounds the interpretations of such regressions and that analyses
of the role of epistasis in adaptation may be meaningful only when the actual
ordered sequence of substitutions is known.

We performed the same kinds of regressions as Chou et al. and Khan et al.
\cite{Chou:2011zr, Khan:2011ly} on adaptive walks simulated on NK landscapes.
Specifically, we computed rank regression coefficients of background fitness
versus fitness effect for the first five substitutions in such adaptive walks (see
Methods).  The distribution of average regression coefficients in Fig.~5(a) shows
a bias toward negative values similar to those seen in bacterial experiments
\cite{Chou:2011zr, Khan:2011ly}, suggesting that epistasis becomes more negative
with each substitution and decelerates the pace of adaptation. However, this
interpretation is contradicted by our results above: Fig.~3 clearly shows that, on
average, epistasis becomes more positive with each substitution. Figs.~1,2, and 4
also support this view: epistasis is initially disruptive to the large fitness
gains of early adaptive changes, but then facilitates later adaptive steps. And,
finally, Fig. 5(b) shows that a genotype along the line of descent is typically
more fit than would be predicted from the fitness effects of its component
mutations in the ancestor, and that this synergistic effect increases along
adaptive walks.

Our results imply that that regression analysis of fitness effects on different
genetic backgrounds (e.g. those performed by Khan et al.  \cite{Khan:2011ly} and
Chou et al. \cite{Chou:2011zr}) may be misleading. To clarify this issue, we
examined the mean regression coefficients for those adaptive walks that were
unequivocally accelerated by epistasis -- namely, those walks in which the effect
of each subsequent mutation was greater than expected under multiplicativity. Even
when restricted to these highly synergistic walks, the regression analyses of the
type shown in Fig.~5a.  are most often negative and so would erroneously suggest
increasing antagonism (Supplemental Figure 13).  Furthermore, we also performed
random (neutral) walks, in which substitutions along the line of descent are
equally likely to exceed or fall short of their expected multiplicative fitness,
given the fitness effect in the ancestor; even in these walks, regression
coefficients of the type studied by Khan et al. \cite{Khan:2011ly} and Chou et al.
\cite{Chou:2011zr} tend be negative (Supplemental Figure 14).  These data confirm
that regressions of fitness effects against the fitnesses of genetic backgrounds
cannot be reliably used to infer whether epistasis has slowed or accelerated
epistasis, at least in the NK model. 

The tendency of these regressions toward negative slopes may be caused by the
well-known confound of "regression to the mean." In these regressions, the
dependent variable, fitness effect of substitution $i$ on a genetic background, is
mathematically interrelated with the independent variable, the fitness of that same
background. If the fitness of the background genotype is very poor, then the epistatic effects of
its alleles are statistically likely to be unusually poor. Any change that perturbs these
epistatic effects is likely to improve their fitness contributions. Therefore, a
substitution in a very unfit background is likely to show a large beneficial
effect simply by perturbing the fitness effects of interacting sites. A similar
argument can be made to explain why fitness effects are often small or even
deleterious in highly fit backgrounds and, by extension, why negative correlations
are a likely consequence of the interdependence between the variables in this
regression. 

The results in this section highlight the difficulty of interpreting experimental
data when the order of substitutions in unknown. The "regression to the mean"
effect we observed may be magnified by the nature of the NK model. Nonetheless,
this example shows that the epistatic properties of a selected sequence of
substitutions may differ strongly and systematically from the broader
pattern of epistasis in the underlying landscape. 

\section{Discussion}

To solve the dual problem of epistasis in evolution experiments -- that epistasis
both shapes the evolutionary process and can be inferred by manipulating the
sites that substitute adaptively -- requires theory beyond current knowledge in
population genetics. Because the term ``epistasis" encompasses all scenarios in which
fitness effects of alleles do not combine independently, no general model of
epistasis has been proposed, let alone analyzed. Instead, exploration of a
few ``toy" models has led to appreciation of the subtle and significant
ways that epistasis complicates our understanding of evolution. 

Here we have used the NK model to contravene the intuitive notion \cite{Ostman:2012uq} that
sites substituting one after another will be selected without regard to their
epistatic interactions. To summarize, we have shown that, even when mutations are
rare, evolution selects among possible substitutions based upon the number and
direction of connections to other loci, and that these selective biases change
substantially along the course of adaptation. In the NK landscapes, epistasis is
less prevalent then the random expectation early in adaptation and much more
prevalent later, with a concomitant shift from predominantly antagonistic
interactions early in adaptation to synergistic and sign epistasis later in
adaptation. Additionally, sites with more epistatic influences on other loci are
more likely to substitute early than late in adaptive evolution. These results
suggest that even the most basic evolutionary process acting in the context of a
simple fitness landscape can produce a complex expectation for epistasis.
Experimentalists must account for this baseline action of natural selection on
epistasis among substitutions if they hope to infer the properties of the fitness
landscapes underlying experimental or natural evolutionary outcomes. 

The basic intuition behind our results is simple. Early on, large-effect mutations
tend to act antagonistically, which suppresses the frequency of epistasis among
subsequent substitutions. Later on, the only way to achieve further fitness gains
is by fortunate sign epistasis, and this effect tends to augment the appearance of
epistasis as the population approaches a fitness peak.

We have focused on evolution by sequential fixation of beneficial substitutions in
order to demonstrate that this seemingly simple case conceals several layers of
complexity. However, our results suggest patterns, such as a decrease in observed
epistasis among early substitutions and an increase in epistasis among later ones,
that extend to polymorphic populations as well. By focusing on the differences
between the distribution of epistasis among all sites, and the specific sequence
of substitutions, our approach highlights the potential for misleading inferences
from evolution experiments when the order of substitutions is unknown.
Specifically, it may be difficult to reliably determine how the sign of epistasis
varies with fitness using only the fitnesses of an ancestral, derived, and
possible intermediate genotypes. Regression analyses that ignore substitution
order might suggest that epistasis is decelerating adaptation (\cite{Chou:2011zr,
Khan:2011ly}, Fig.~5), whereas in fact epistasis has had an accelerating effect on
the trajectory of fitnesses along the actual path of adaptation. Indeed, our
reanalysis of data from \cite{Khan:2011ly} supports this possibility. Such
discrepancies illustrate the importance of measuring the order in which
substitutions occur, in future experimental studies, in order to understand how
epistasis has shaped a population's trajectory.

While the NK model has the advantages of a tunable level of epistasis and an
extensive history of prior work, it certainly does not capture the full range of
possibilities of interactions among genes. Our results with larger numbers of
alleles in the NK model, with a computational model of RNA folding, and with data
from evolving microbial populations, suggest that the patterns we have identified
in simple cases may be even more pronounced in models that better approximate
biological complexity. However, the most important conclusion from our study is
that new methods of data-gathering, such as cheap whole-genome sequencing and
high-throughput fitness assays, will not suffice to answer basic questions about
the role of genotype-phenotype maps in evolution.  Pioneering studies have
provided compelling examples of the ubiquity of epistasis, but they also serve to
exemplify the substantial gap that separates data from hypotheses in experimental
evolution. To definitively link gene interactions to the rate or predictability of
adaptation will require a significant expansion of the theory of population
genetics, and a vital first step is serious engagement with "toy" models of
interacting loci.

\section{Methods}

\subsection{Mathematical Fitness Landscapes}

Invented by Stuart Kauffman to describe evolution on rugged fitness landscapes
\cite{Kauffman:1987bq,KAUFFMAN:1989kx,Kauffman:1993fk}, the NK model produces
complex but computationally tractable genotype-fitness maps using only the
parameters $N$, $K$, and $A$. The parameter $N$ defines the number of sites, each
of which can assume any of $A$ alleles. The fitness of a genotype is calculated in
two steps: first, the fitness contribution of each site is  determined by
reference to a table of pre-calculated values; second, these fitness contributions
are multiplied together and the N-th root of this product is taken as fitness.
When $K$ is zero, the fitness contribution of a site depends only on its own
allele, and not on the state of other sites; the lookup table for each site
therefore contains only $A$ possible fitness contributions. When $K > 0$, the
fitness contribution of a site depends on its own allele as well as the alleles at $K$ other sites,
yielding a lookup table with $A^{K+1}$ entries.
By specifying that each of the entries in the lookup tables for all $N$
sites are drawn independently from a broad distribution (in our case the uniform
distribution), the NK model ensures that the fitness effect of a substitution
depends strongly and randomly on some fraction of the genetic background, determined
by $K$. $K$ is constant across sites and genotypes for a particular landscape, and
the $K$ sites upon which each locus depend are drawn uniformly from the $N-1$
possibilities. 

Gene interactions described by the NK model are directional: the fitness
contribution of site $i$ may depend on the state of site $j$ without implying that
the contribution of $j$ also depends on $i$. We can therefore categorize
relationships between sites: $i$ is \emph{downstream} of $j$ if the fitness
contribution of $i$ depends on $j$, and $j$ is conversely \emph{upstream} of $i$
in this example. While this direction of influence is significant for some of our
results below, we note a subtle confusion between epistasis as defined in the NK
model, and epistasis as defined by an experimentalist. This confusion stems from
the fact that an experimentalist measures the fitness effect of a substitution,
while the NK model considers fitness contributions of sites. Therefore, the
fitness effect of a substitution at site $i$ will show epistasis with site $j$ not
only if $i$ is downstream of $j$, but also if it is upstream of $j$. Similarly,
the fitness effect of a substitution at $i$ will be epistatic with $j$ if both $i$ and $j$
are upstream of some mutual site $k$, even if neither $i$ nor $j$ directly
influence one another. To minimize this potential confusion, we follow the
empirical definition and use the term `epistasis' to refer to any case in which
the fitness effect of a substitution at a site depends on the state of some other
site. This usage makes the common assumption that independent fitness effects are
multiplicative.

Aside from NK landscapes, we also explore epistasis and evolutionary dynamics using
RNA folding landscapes, as well as experimental bacterial data, described below.

\subsection{Evolutionary Simulations}

To investigate epistasis among genetic substitutions, we employed two types of
simulations: adaptive walks, which model simplified fixation dynamics in
essentially monomorphic populations; and individual-based Monte Carlo simulations of
potentially diverse populations. In an adaptive walk the population is represented
by a single genotype, and a mutation to any of the $A-1$ alternative alleles at
any site is a candidate substitution. For simplicity, we assume that the
probability of fixation for a mutation is directly proportional to its selective
coefficient, $s_i = \frac{w_i}{w}-1$ where $w$ is the fitness of the currently
fixed genotype and $w_i$ the fitness of the mutant $i$. Evolution is then a
Markov process with transition probabilities defined by

\begin{equation}
P_{i} = \frac{s_i}{\sum \limits_{i \in M} s_i}
\end{equation}

where $M_i$ is the set of all adaptive, one-mutant neighbors of the current genotype
\cite{GILLESPIE:1984uq,Orr:2002vn}. These walks therefore substitute one mutation 
at a time, strictly increasing in fitness until a local maximum is reached.

We used the Wright-Fisher to describe evolution in polymorphic populations:
an asexual population of fixed size $n$ reproduces with discrete generations and selection on
fertility. Mutations occur at Poisson frequencies according to the per-genome rate
$\mu$. Because our goal is to investigate evolutionary dynamics when $n\mu =
\theta$ is near or greater than 1, we required a method of detecting substitutions
that does not depend on independent, well-demarcated fixation events. We therefore
trace the line of descent from the final population back to the initial
generation and record changes along this lineage. To minimize the false
identification of polymorphisms as substitutions, we ran these simulations for $n$
generations past the desired ending time-point. We then selected the most fit
individual and traced its lineage back to the start, discarding any genetic
changes that arose during those last $n$ generations.

To compare our results with empirical data, we performed two types of regressions
on the fitness effects of the first few substitutions in 
evolutionary simulations. Following recent experimental examples
\cite{Chou:2011zr, Khan:2011ly}, we examined the first five substitutions in a
simulation and measured the fitness effect of each of these substitutions with all
combinations of the allele states at the other four sites. The fitness of each of
the sixteen genetic backgrounds is taken as the independent variable, and the
fitness effect of the focal substitution as the dependent variable; a separate
regression was performed for each of the five sites, though the resulting five
correlation coefficients are not independent. These analyses were contrasted with
a second type of regression, in which the ranks of epistatic deviations of each
successive substitution were compared to their order of substitution. If we
consider the first five substitutions, then there are four epistatic deviations:

\begin{eqnarray}
e_{12} &=& W_{12} - W_1 W_2\\ 
e_{123} &=& W_{123} - W_1 W_2 W_3\\ 
e_{1234} &=& W_{1234} - W_1 W_2 W_3 W_4\\ 
e_{12345} &=& W_{12345} - W_1 W_2 W_3 W_4 W_5
\end{eqnarray}

where $W_{12}$, for example, represents the fitness of the genotype with the first
two substitutions, divided by the fitness of the ancestor. Because these
regressions are based on four points, we use this analysis only to classify
epistasis along a walk; for example, we highlight the significance of walks in
which $e_{12345} > e_{1234} > e_{123} > e_{12}$ as examples where epistasis
consistently leads to greater than expected fitness.

In both cases, we ignore simulations with little epistasis among the first five
substitutions in order to avoid spurious correlations; our filtering removed about
20\% of simulations. Because we remove walks without regard to the direction of
epistatic effects, this removal should not bias any results.

New landscapes were generated for each replicate simulation. In both types of
populations, a simulation began from a randomly drawn genotype with fitness in a
certain range, usually close to the 50th percentile (that is, within 0.002 of the
desired starting fitness).

\subsection{RNA Fitness Landscapes}

We also performed Wright-Fisher simulations of evolving RNA populations using the
Vienna RNA folding package, version 1.8.5, with default folding parameters. RNA
sequences of 72 bases in length constituted the genotypes, and the predicted
minimum-free-energy structures determined the phenotypes. Fitness of an RNA
genotype was calculated as $(1 + s)^{-d}$ where $d$ denotes the tree edit distance
between the RNA's phenotype and a defined optimal phenotype. Here $s$ quantifies
the strength of selection and is equivalent to the multiplicative selective
coefficient associated with a mutation that changes $d$ by a single unit; $s =
0.01$ in the results shown.  The tree edit distance algorithm, included in the
Vienna package, determines the minimum number of steps from a group of edit
operations that are needed to transform one structure into another.  The initial
genotype was drawn randomly, and the optimum phenotype, used to impose directional
selection, was also created by randomly drawing genotypes and discarding those
whose minimum free energy structure is the trivial, unfolded state. This optimum
was also required to be 40 units from the phenotype of the initial genotype so
that the pressure to adapt was strong and uniform across replicates. Simulations
were run for 50,000 generations.

Substitutions in RNA simulations were determined by tracing a line of descent, 
as described above for NK simulations. Because our goal was to study
adaptation among beneficial mutations, we filtered the resulting records of
substitutions by ignoring adjacent pairs on the line of descent when both members
of the pair were neutral or deleterious on the background in which they originally
fixed. In practice, less than 1\% of pairs were excluded by this rule, so our
results are not sensitive to this criterion.

\clearpage

\begin{figure}[!htb] \centering  \includegraphics[width=15cm]{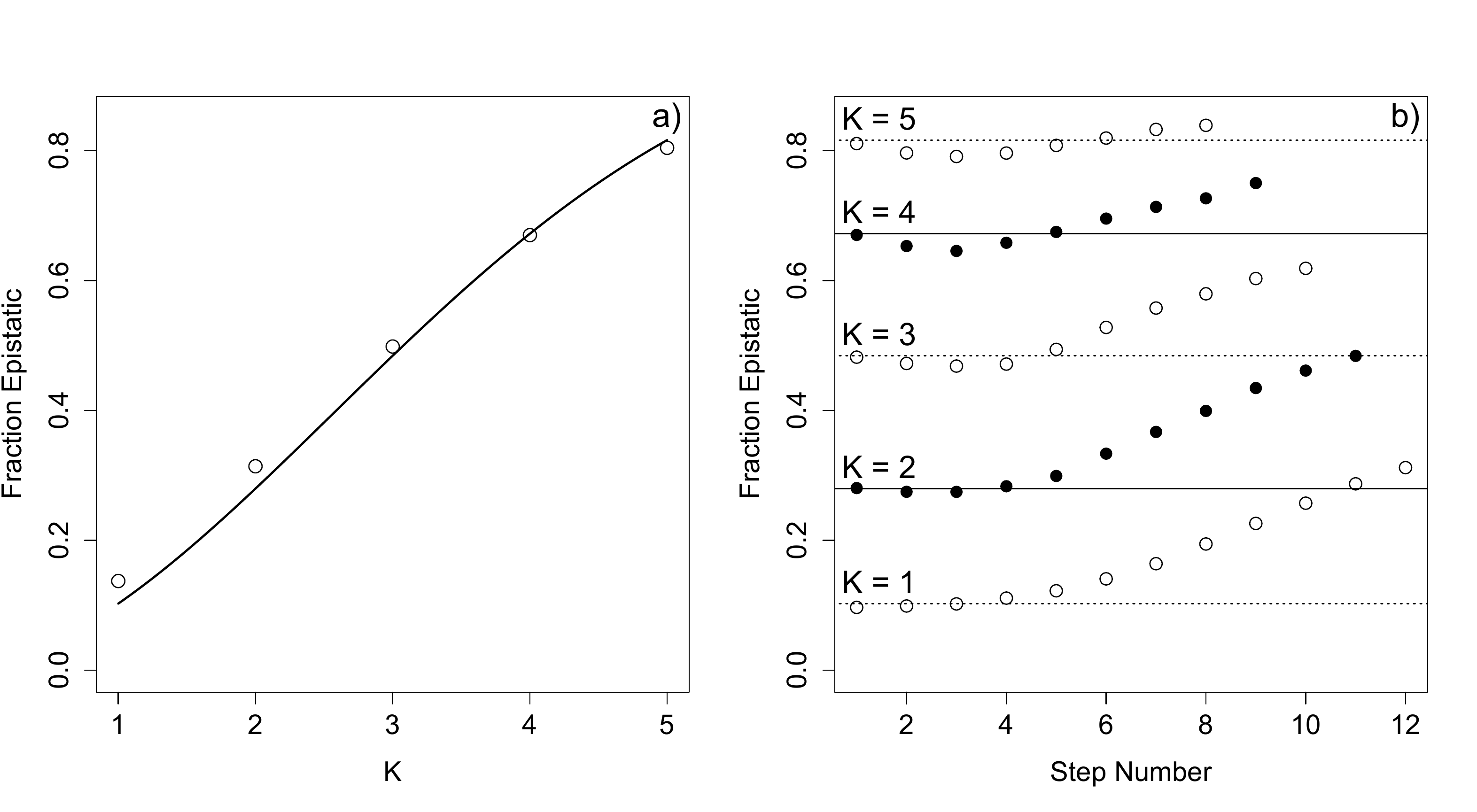}
\caption{\emph{a)} The overall amount of epistasis along an adaptive walk in an NK
landscape roughly agrees with its random expectation. The dots show the frequency
of epistatic interactions between substitution $i$ and its immediate successor
$j$, averaged across all steps in the adaptive walk. The solid line depicts the
predicted incidence of epistasis, if sites are chosen to substitute randomly
(see supplement). $N = 20$ and $A = 2$. Standard errors
are less than 0.001. \emph{b)} The frequency of epistasis between subsequent
substitutions is depressed compared to the random expectations, early in
adaptation, and augmented late in adaptation.  Dots show the frequencies of
epistasis between substitution $i$ and its immediate successor $j$, indexed by the
position of substitution $i$ along the adaptive walk. Lines indicate the
corresponding random expectation. Standard errors are less than 0.01 for all
plotted means.} \label{fig:K1} \end{figure}

\begin{figure}[!htb] \centering \includegraphics[width=16cm]{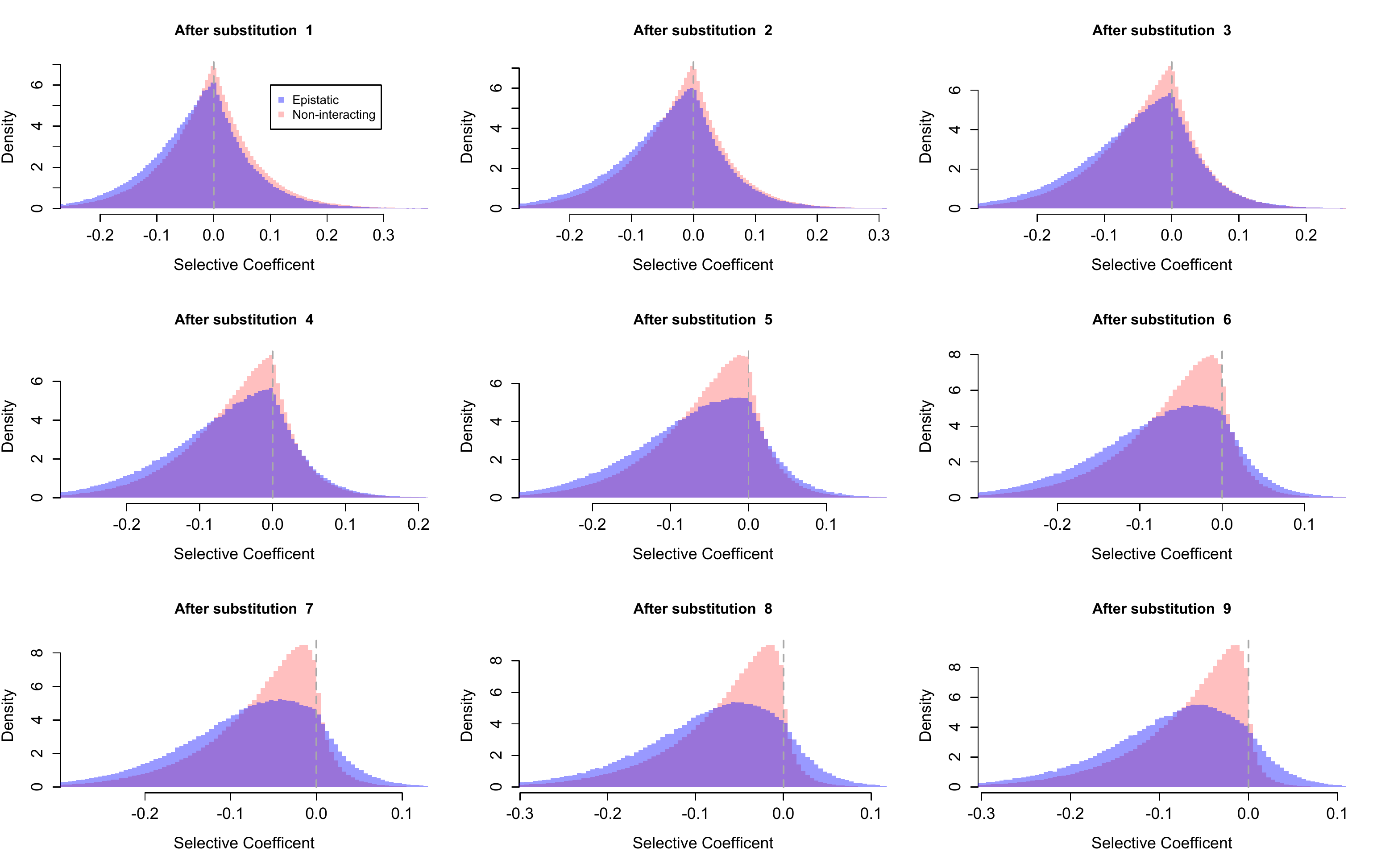}
\caption{Mutations at sites that interact with recent substitutions are less
favorable early in evolution and more favorable late, in comparison to
non-interacting sites. Histograms depict the frequencies of the selective
coefficients of mutations across many replicate adaptive walks with parameters $N
= 20$, $K = 1$ and $A = 2$. After an adaptive substitution occurs at site $i$, all
$(A-1)(N-1)$ other possible mutations are inspected and classified by whether
their fitness effects depend on site $i$ epistatically, or are independent of it.
As the adaptive walk proceeds, the selection coefficients of available mutations
shift towards negative values in general. At the same time, among the adaptive
mutations available late in evolution, the great majority of them interact with a
recent substitution. Thus, epistasis tends to retard the substitution of interacting
sites early in evolution, and promote such substitutions late in evolution.
}
\label{fig:K1} \end{figure}

\begin{figure}[!htb] \centering \includegraphics[width=16cm]{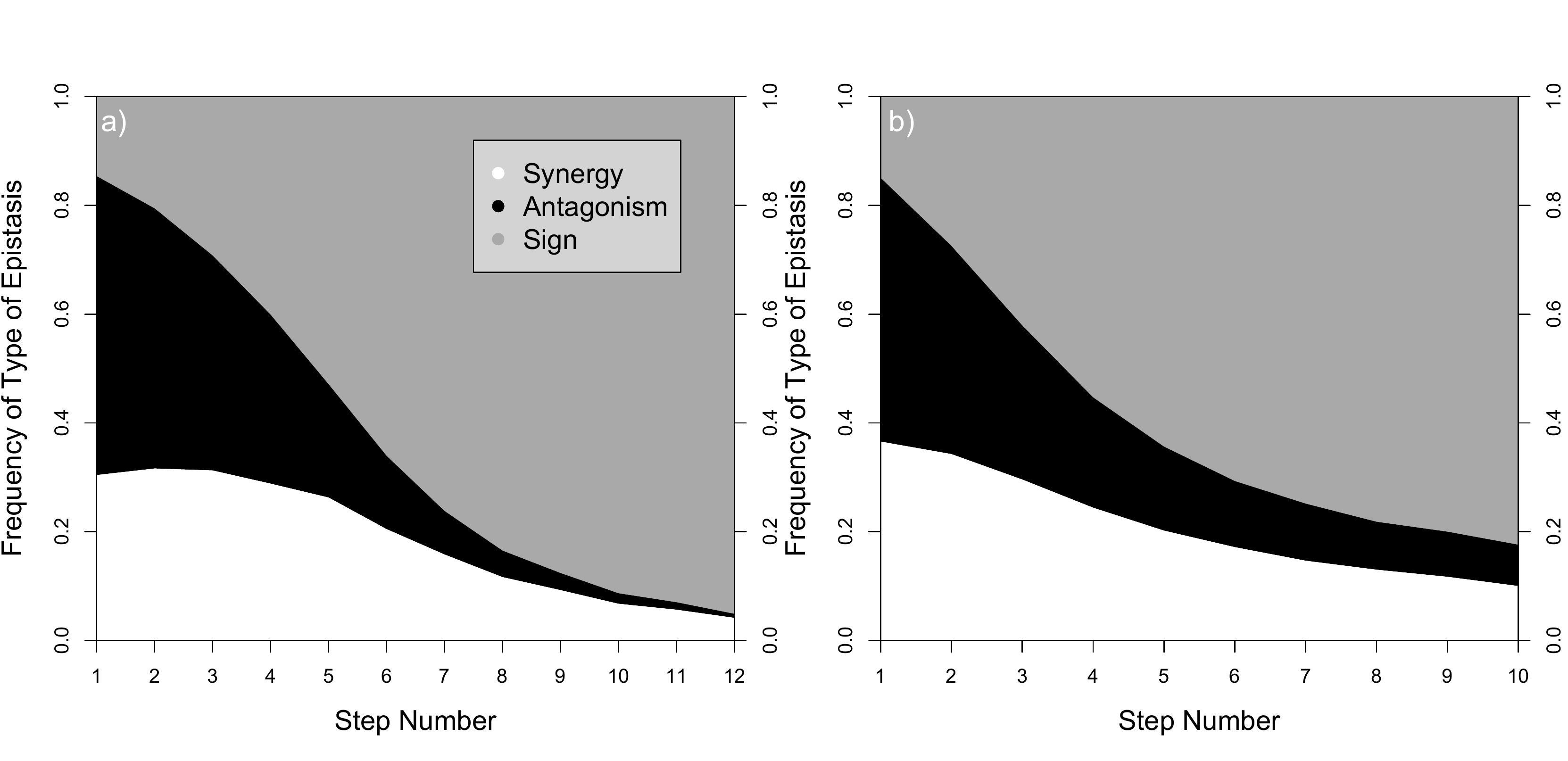}
  \caption{Prevalence of
synergy, antagonism, and sign epistasis among pairs of consecutive substitutions
$(i,j)$ along adaptive walks on $NK$ landscapes. Such walks are
characterized by an abundance of antagonism early in adaptation, and an abundance
of sign epistasis late in adaptation. a) $K = 1$, b) $K = 5$.  $N = 20$ and $A =
2$.} \label{fig:K1} \end{figure}

\begin{figure}[!htb] \centering  \includegraphics[width=14cm]{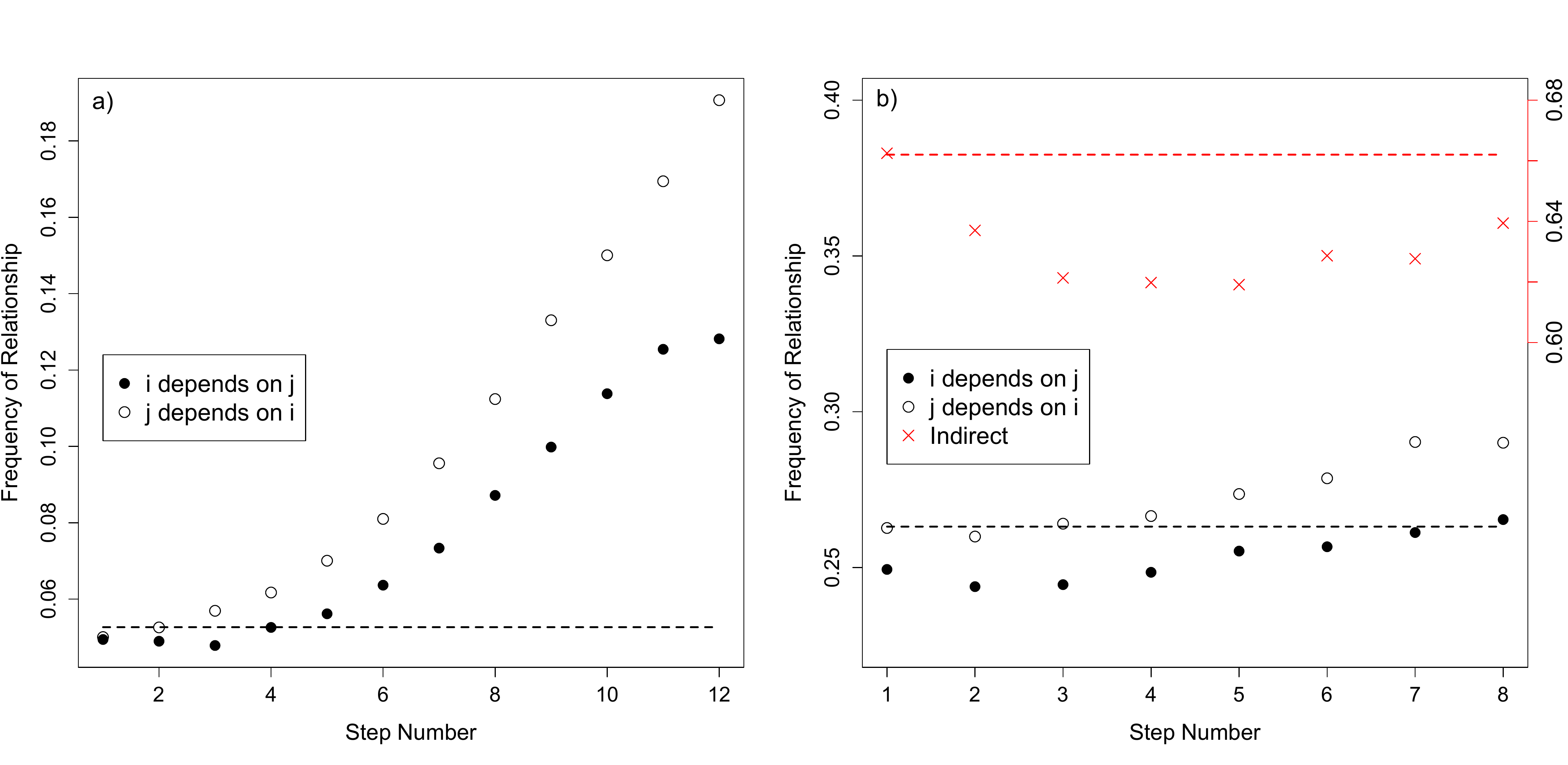}
\caption{Observed frequency of directional epistatic interactions between
substitution $i$ and its immediate successor $j$ along adaptive walks on $NK$
landscapes. a) $K = 1$; b) $K = 5$.  A single pair may be counted as more than one
type of epistasis, as in the case of reciprocal interactions.  The dashed lines
depict the predicted incidences if substitutions are chosen randomly. For $K > 1$, another class of
epistasis is possible: both $i$ and $j$ may jointly influence a third site, which
we refer to as an indirect interaction. $N = 20$ and $A = 2$. Standard errors are
less than 0.01 for all plotted means.} \label{fig:K1} \end{figure}

\begin{figure}[!htb] \centering   \includegraphics[width=14cm]{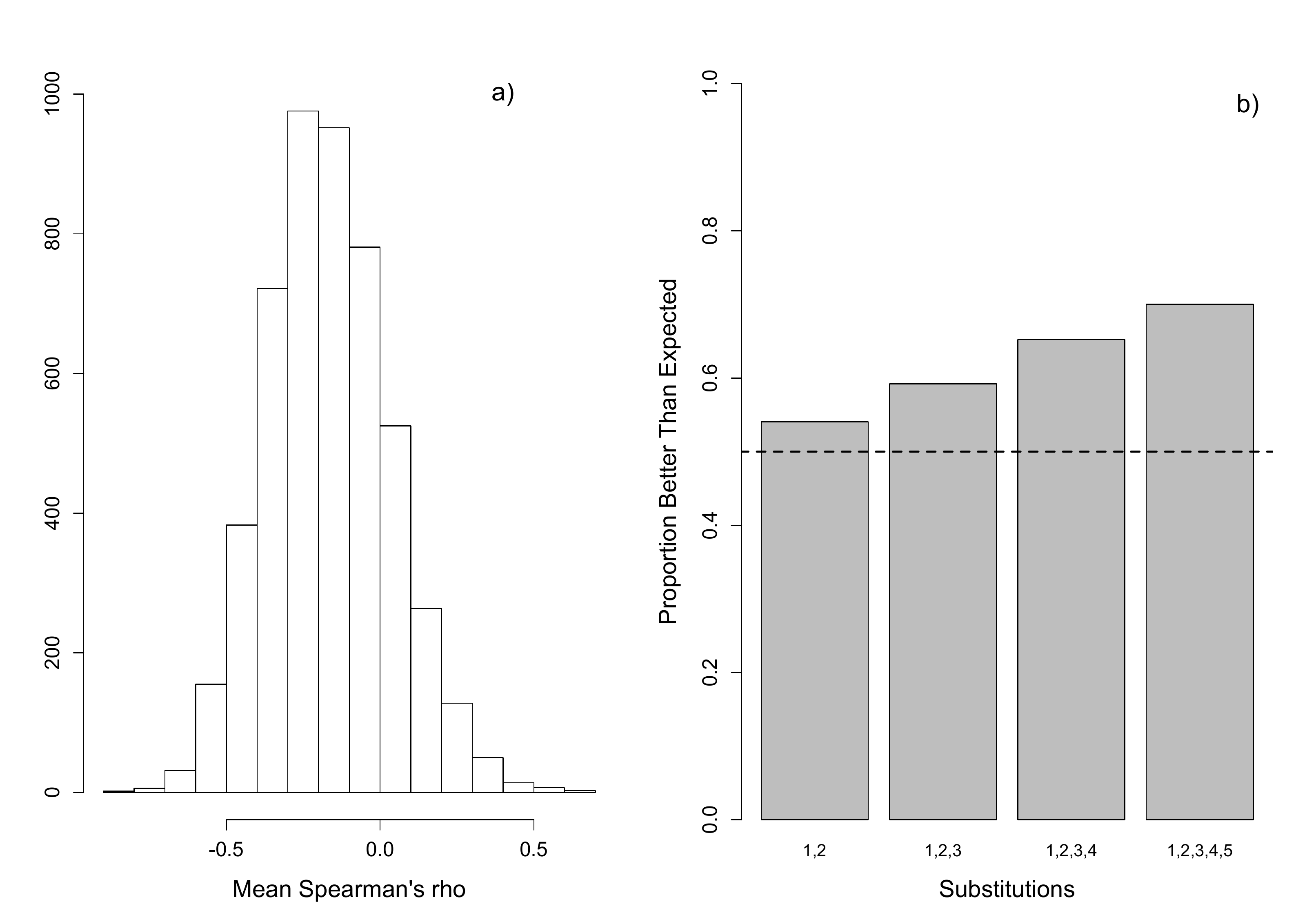}
\caption{Two views of epistasis among the first five substitutions in an adaptive walk. a) Spearman
regression coefficients for the relationships between the fitness effect of a
substitution and the fitness of the genetic background in which that substitution
is made \cite{Chou:2011zr,
Khan:2011ly}. Each substitution is tested against the sixteen genetic backgrounds
comprising all combinations of the other four substitutions. b) Proportion of
combinations of substitutions along the line of descent which exceed the fitness
expected from the fitness effects of the individual mutations in the ancestor.
Expectations are computed according to Equations 7-10. Standard errors are less
than 1\%. In both figures, replicates are filtered to remove adaptive walks with
too little epistasis among substitutions (see methods) -- data shown in panel (b) are additionally filtered to remove cases where $W_{12} = W_1 W_2/W$. 
The analysis on
the left panel would suggest that epistasis is primarily antagonistic, whereas in
fact the right panel shows that synergistic interactions dominate along the walk. $N = 20$, $K
= 5$ and $A = 2$.} \label{fig:K1} \end{figure}

\clearpage

\section{Appendix}

\subsection{Neutral Prevalence of Epistasis in the NK Model}
The simplicity of the NK model allows the probabilities of different epistatic 
relationships between substituting sites to be calculated, assuming that sites 
are chosen to substitute at random.
For distinct loci $i$ and $j$, the random (neutral) walk prevalences are given by:

\begin{eqnarray}
\label{eq:neutral_expectation}
p_{N,u} &=& \frac{K}{N-1} \\
p_{N,d} &=& \frac{K}{N-1} \\
p_{N,i} &=& 1 - \left(1 - \frac{K}{N-1}\frac{K-1}{N-2}\right)^{N-2} \\
p_{N,m} &=& (1 - p_{N,u}) (1 - p_{N,d}) (1 - p_{N,i})\nonumber\\
 &=& \left(1 - \frac{K}{N-1}\right)^2 \left(1 - \frac{K}{N-1}\frac{K-1}{N-2}\right)^{N-2}
\end{eqnarray}

\subsection{Estimating Substitution Probabilities for Upstream and Downstream Epistatic Pairs}

Our results focus on two main types of epistasis: a site may be upstream or downstream of another site. While both types of influence create an epistatic interaction between the fitness effects of changes at both sites, simulation results reveal a preference for interactions where the first site to substitute, $i$, is upstream of $j$. Here we try to quantify both hypothesized effects of epistasis -- the detrimental undermining of fixed beneficial effects, and the positive potential for epistasis to uncover new directions for adaptation -- by calculating conditional probabilities in the NK model.

The first goal is to calculate the probability that a site $j$ with alleles $b$ and $B$ is capable of a beneficial substitution after a beneficial change at site $i$ from allele $a$ to allele $A$, assuming epistasis between $i$ and $j$. We consider two ways in which sites $i$ and $j$ can interact: $i$ is \emph{upstream} of $j$, and so $i$ determines the effect of substitution at $j$, or $i$ may be \emph{downstream} of $j$; we will ignore the case where each site may \emph{mutually} influence one another. In both cases, we want to solve:

\begin{equation}
\label{inequal}
P\{W_{AB} > W_{Ab}\} = P\{w_A(B) w_B(A) > w_A(b) w_b(A) \}
\end{equation}

\noindent where the notation $w_A(b)$ indicates the fitness effect of allele $b$ at site $j$ when site $i$ has allele $A$, and $W_{AB}$ represents the total fitness contribution of both sites when site $i$ has allele $A$ and site $j$ has allele $B$. Because fitness is a product across sites, we define $W_{AB} = w_A(B) w_B(A)$. While fitness is technically the Nth root of the product across all $N$ sites, our results below depend only on the ranks of fitness contributions; we therefore ignore this complication. 

These calculations, performed in detail below, demonstrate that the chance for a site to be evolvable -- selectively favored to substitute -- is diminished if that site's epistatic partner has beneficially substituted, and that this reduced probability is caused by the fact that such a substitution will disrupt the genetic background that was favorable to the prior substitution. Strikingly, these costs to epistasis seem to be consistent between the two types of epistasis, despite the observed preference for interactions where $i$ is upstream of $j$ (Figure 3). While these two kinds of epistasis are equally likely to undermine the benefit of previous substitutions, and so have a net deleterious effect, they differ in a more subtle aspect of evolvability. Natural selection should exhaust the set of potentially beneficial changes and so late in adaptation we expect evolvable sites to be rare. To simulate late adaptation, we could assume that $w_a(B) w_B(a) < w_a(b) w_b(a)$ -- that is, that site $j$ was not evolvable prior to change at site $i$. Does this assumption reveal a difference between the two classes of epistasis?

Simple numerics demonstrate that this additional assumption favors substitution at $j$ when $j$ is downstream of $i$: 35.8\% $\pm$ 0.15\% of such sites are favorable after substitution at $i$, as opposed to only 14.4\% $\pm$ 0.19\% when $j$ is upstream of $i$. While these probabilities could be derived as above, a more intuitive approach is to consider the effect of each assumption of the fitness effect of a substitution at site $j$, after substitution at site $i$. This effect is related to the ratio:

\[
\frac{w_B(A) w_A(B)}{w_b(A) w_A(b)}
\]

In either model, our expectation of the value of the denominator is increased by the knowledge that $w_a(B) w_B(a) < w_a(b) w_b(a) < w_A(b) w_b(A)$. When $j$ is downstream of $i$, this increase is partially mitigated by two factors: $w_B(A) = w_b(A)$, and the remaining term in the numerator, $w_A(B)$, is an independent draw. However, when $j$ is upstream of $i$, then $w_B(A)$ is an independent draw, and $w_A(B)$ is expected to be smaller than an independent draw, since it equals $w_a(B)$. This asymmetry between the evolutionary acceptability of two directions of epistatic influence is therefore caused by another asymmetry: the high likelihood that sites are unevolvable later in an adaptive process. 

\subsection{Site $i$ is upstream of $j$}
When $j$ depends on $i$, but $i$ is independent of $j$, then we have $w_b(A) = w_B(A)$ and $w_b(a) = w_B(a)$. Eq. \ref{inequal} therefore reduces to:

\[
P\{W_{AB} > W_{Ab}\} = P\{w_A(B) w_B(A) > w_A(b) w_b(A)\} = P\{w_A(B) > w_A(b)\}
\]

Because the beneficial substitution $a \rightarrow A$ occurred on background $b$, and we assume only beneficial substitutions, $w_b(A) w_A(b) > w_b(a) w_a(b)$. We want the distribution of $w_A(b)$, given that:

\[
w_A(b) > \frac{w_b(a) w_a(b)}{w_b(A)}
\]

All four of these variables are standard random uniform variates, and we assume that they are independent. Let $x = w_b(a)$, $y = w_a(b)$, $z = xy$, and $v = w_b(A)$. We can therefore solve for the convolution $z = xy$, then for the quotient distribution $\frac{z}{v}$.

We can find the distribution of $z$ by constructing the cumulative distribution function:

\[
F_{xy}(z) = z + \int_z^1 \frac{z}{x} dx = z(1 +\left. ln x \right |_z^1) = z(1 - ln z)
\]

Then we differentiate to obtain the p.d.f.:

\[
f_{xy}(z) = -ln(z)
\]

Next, we obtain the convolution of $z$ over $v$ by a similar construction. Let $q = \frac{z}{v}$, and note that $q$ ranges from zero to infinity. A simple geometric approach reveals that $q$ is piecewise at 1. When $q \leq 1$, we can define the c.d.f.:

\[
F_{z/v}(q | q \leq 1) = \int_0^q -ln (z) \left(1 - \frac{z}{q}\right) dz 
\]

\[
f_{z/v}(q | q \leq 1) = \frac{1}{4} (3 - 2 (ln (q) + 1)) 
\]

\noindent and when $z > 1$:

\[
F_{z/v}(q | q > 1) = \int_0^1 -ln (z) \left(1 - \frac{z}{q}\right) dz 
\]

\[
f_{z/v}(q | q > 1) = \frac{1}{4q^2}
\]

However, our goal is to find the distribution of $w_A(b)$ given that it is greater than $q$, and since $0 \leq w_A(b) \leq 1$, this latter piece of $f(q)$ is not needed.

\begin{eqnarray*}
f(x = w_A(b)) &=& \frac{\int_0^x  3 - 2 (ln (q) + 1) dq}{\int_0^1 dx \int_0^x  3 - 2 (ln (q) + 1) dq} \\
&=& \frac{x(3 - 2 ln(x))}{2}
\end{eqnarray*}

Focal site $j$ will be beneficial if $w_A(B) > w_A(b)$, which we can now calculate as:

\begin{eqnarray*}
P\{w_A(B) > w_A(b)\} &=& \frac{\int_0^1 dx \int_0^x y(3 - 2 ln(y)) dy}{\int_0^1 dx \int_0^1 y(3 - 2 ln(y)) dy}\\
&=& \frac{7}{18}
\end{eqnarray*}

\subsection{Site $i$ is downstream of $j$}

If $i$ depends on $j$, but $j$ is independent of $i$, then $w_a(b) = w_A(b)$ and $w_a(B) = w_A(B)$. The condition $w_A(b) w_b(A) > w_a(b) w_b(a)$ therefore simplifies to $w_b(A) > w_b(a)$. Let $x = w_b(A)$ and $y = w_b(a)$ and assume that these are independent uniform random variates. 

\[
p(x | x > y) = 2 \int_0^x dx = 2x
\]

Now we let $v = w_A(b)$ and find the convolution $z = xv | x > y$ by first deriving a c.d.f.

\[
F_{xv | x>y}(z) = 2 \int_0^z x dx + 2 \int_z^1 x \frac{z}{x} dx = z^2 + 2z - 2z^2 = 2z - z^2
\]

\[
f_{xv | x>y}(z) = 2 - 2z
\]

Now, we want to assess the probability that  $w = w_A(B) w_B(A)$, a pair of independent random uniform variates, is greater than $z$. Using the result above that $p(w) = -ln(w)$, we can write this as a ratio of integrals:

\begin{eqnarray*}
P\{w_A(B) w_B(A) > w_A(b) w_b(A)\} &=& \frac{\int_0^1 -ln(w) dw \int_0^w 2 - 2 z dz}{\int_0^1 -ln(w) dw \int_0^1 2 - 2z dz}\\
&=& \frac{7}{18}
\end{eqnarray*}

\clearpage

\subsection{Supplemental Figures}

\begin{figure}[!htb] \centering \includegraphics[width=17cm]{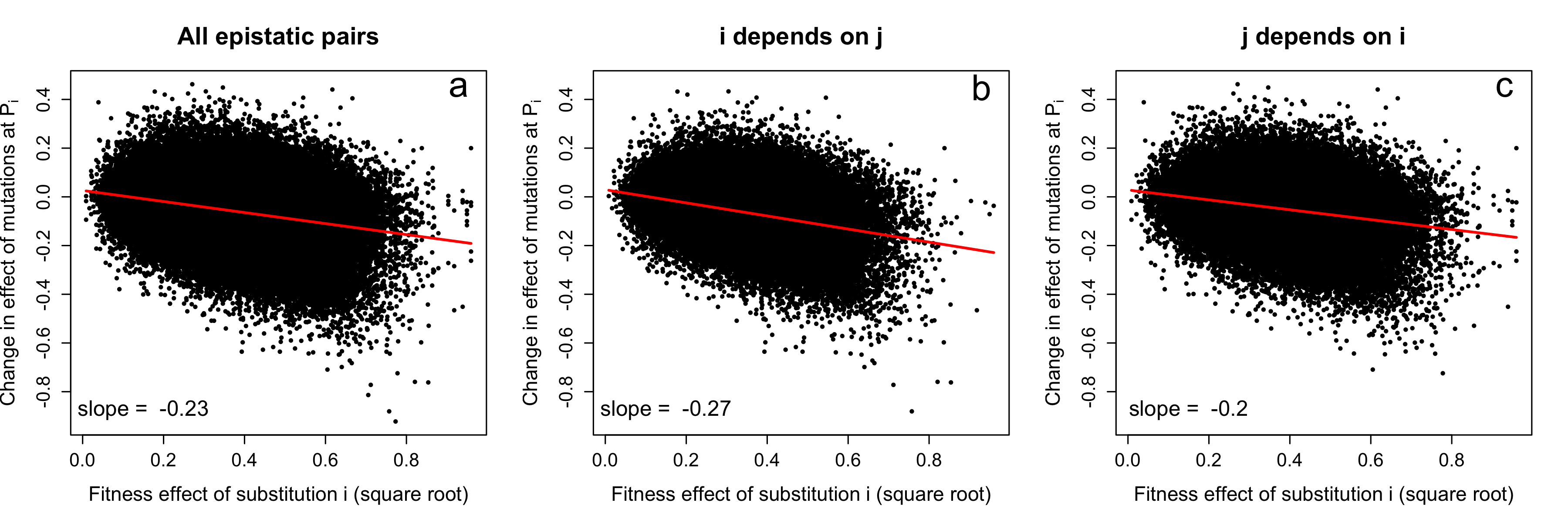} \caption{Relation of the beneficial fitness effect of substitution $i$ to the change in the fitness effect conferred by a mutation at interacting site $j$. Data are drawn from the first step in adaptive walks. a) All types of interactions; b) Interactions in which site $i$ is downstream of site $j$; c) Interactions in which $i$ is upstream of $j$. $N = 20$, $K = 1$ and $A = 2$.} \end{figure}

\begin{figure}[!htb] \centering \includegraphics[width=17cm]{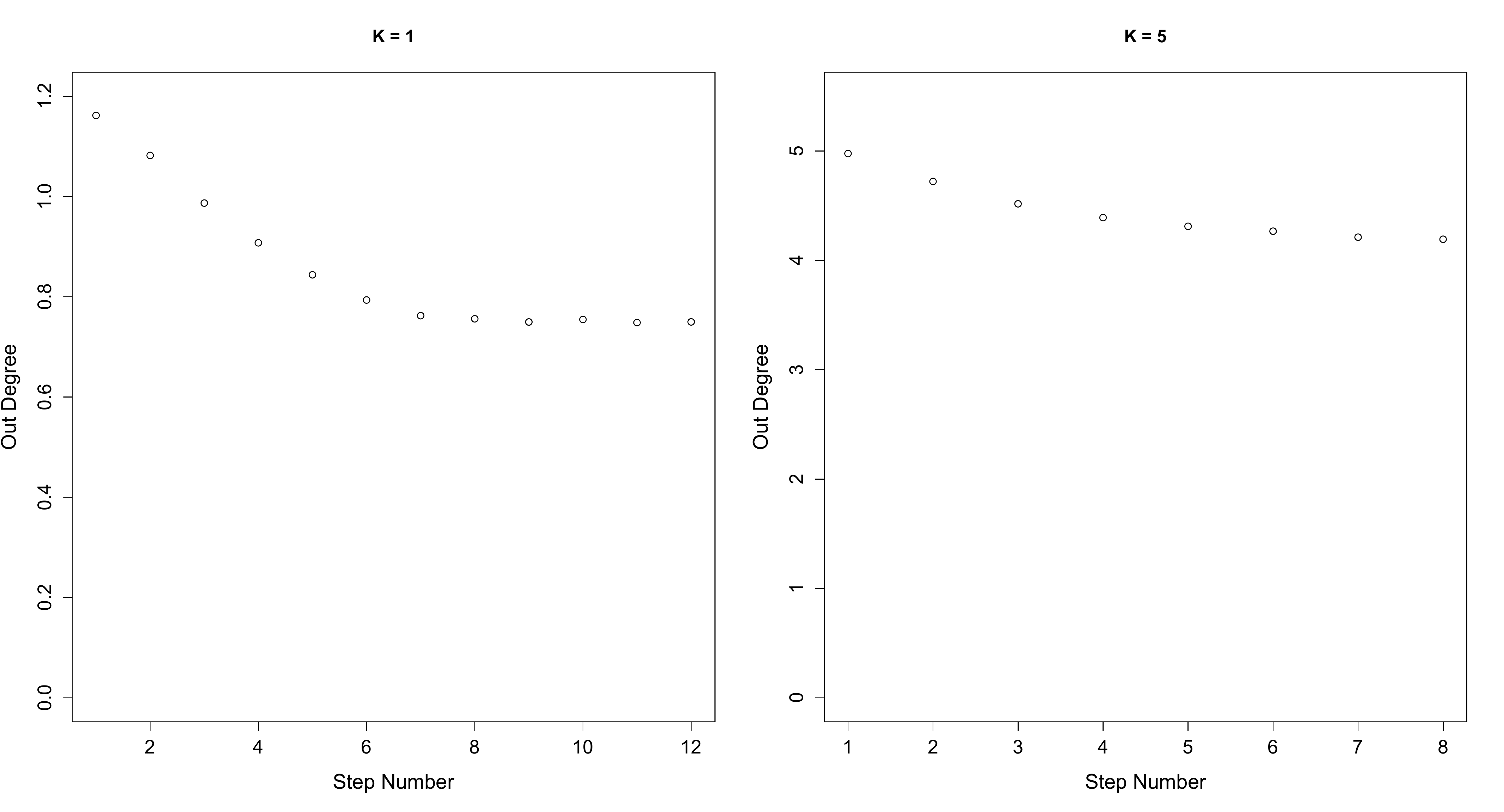} \caption{Mean out-degree (number of loci which depend epistatically on the focal site) of substituted sites. Each plotted point is based on at least 5000 adaptive walks. $N = 20$ and $A = 2$.} \end{figure}

\begin{figure}[!htb] \centering \includegraphics[width=14cm]{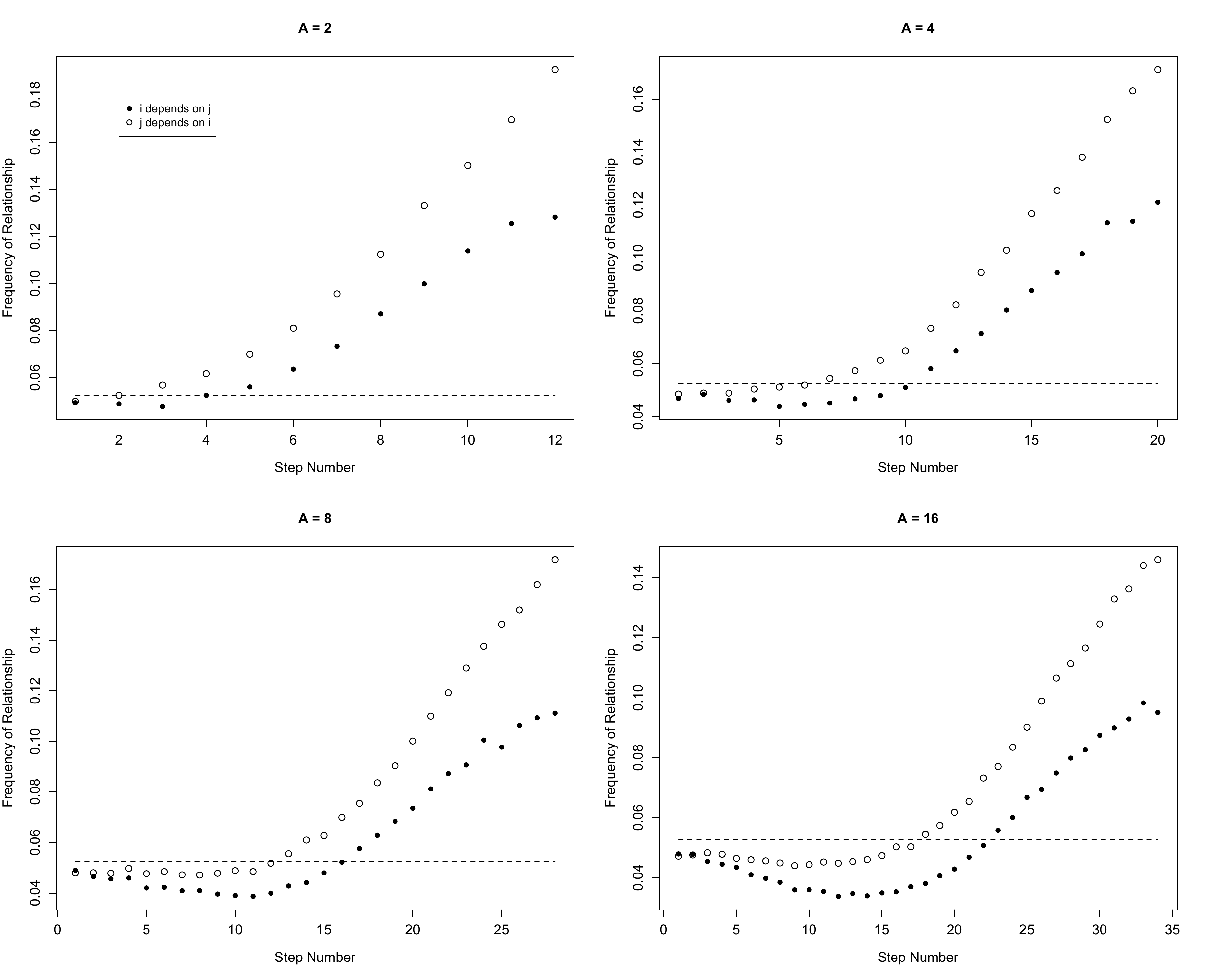} \caption{Observed frequency of directional epistatic interactions between substitution $i$ and its immediate successor $j$ for adaptive walks on $NK$ landscapes with different values of $A$, the number of alleles per locus. A single pair may be counted as more than one type of epistasis, as in the case of reciprocal interactions. The dashed lines depicts the predicted incidences if sites are chosen uniformly at random, as calculated in Eqs.~2\&3. $N = 20$ and $K = 1$. Each plotted point is based on at least 5000 replicates.} \end{figure}

\begin{figure}[!htb] \centering \includegraphics[width=14cm]{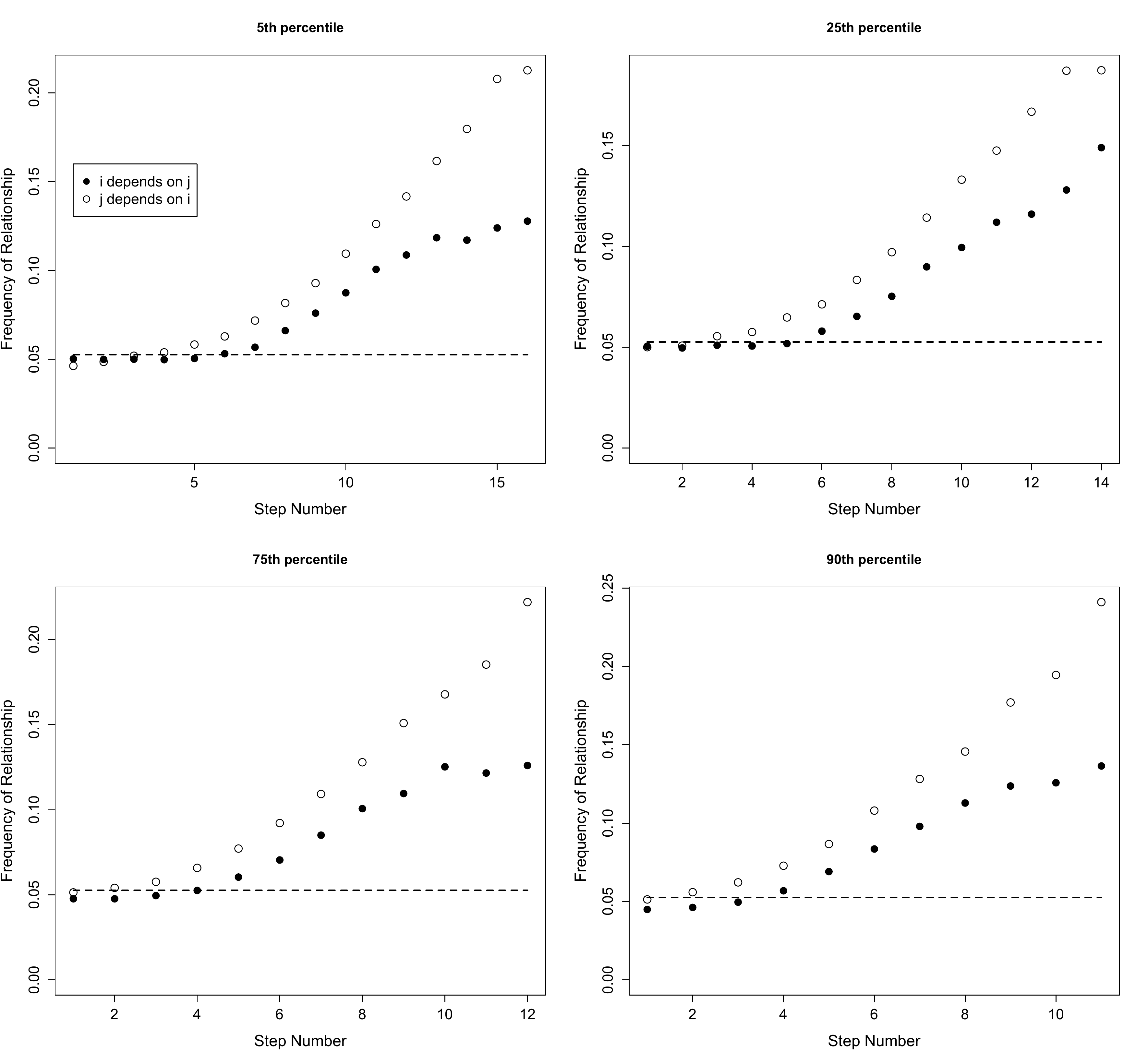} \caption{Observed frequency of directional epistatic interactions between substitution $i$ and its immediate successor $j$ for adaptive walks on $NK$ landscapes for different values of the fitness percentile of the starting genotype. A single pair may be counted as more than one type of epistasis, as in the case of reciprocal interactions. The dashed lines depicts the predicted incidences if sites are chosen uniformly at random, as calculated in Eqs.~2\&3. $N = 20$ and $K = 1$. Each plotted point is based on at least 5000 replicates.} \end{figure}

\begin{figure}[!htb] \centering \includegraphics[width=14cm]{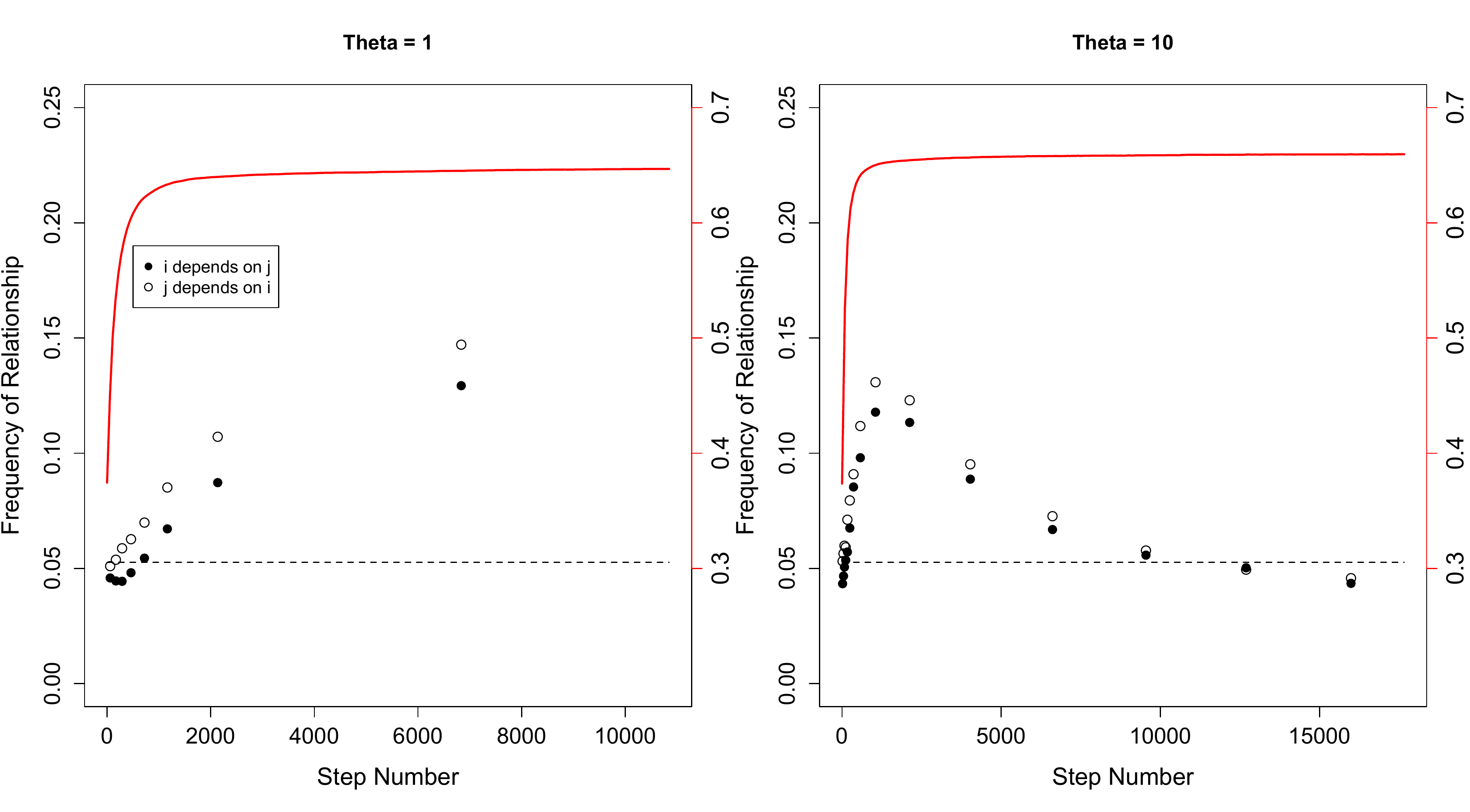} \caption{Observed frequency of directional epistatic interactions between substitution $i$ and its immediate successor $j$ for individual-based simulations with population sizes of 1000 and per-genome mutation rates of 0.001 and 0.01. A single pair may be counted as more than one type of epistasis, as in the case of reciprocal interactions. For $K > 1$, another class of epistasis is possible: both $i$ and $j$ may jointly influence a third site, which we refer to as an indirect interaction. The dashed lines depicts the predicted incidences if sites are chosen uniformly at random, as calculated in Eqs. 2-5. Red lines and corresponding axes depict the mean change in fitness. $N = 20$ and $K = 1$.} \end{figure}

\begin{figure}[!htb] \centering \includegraphics[width=14cm]{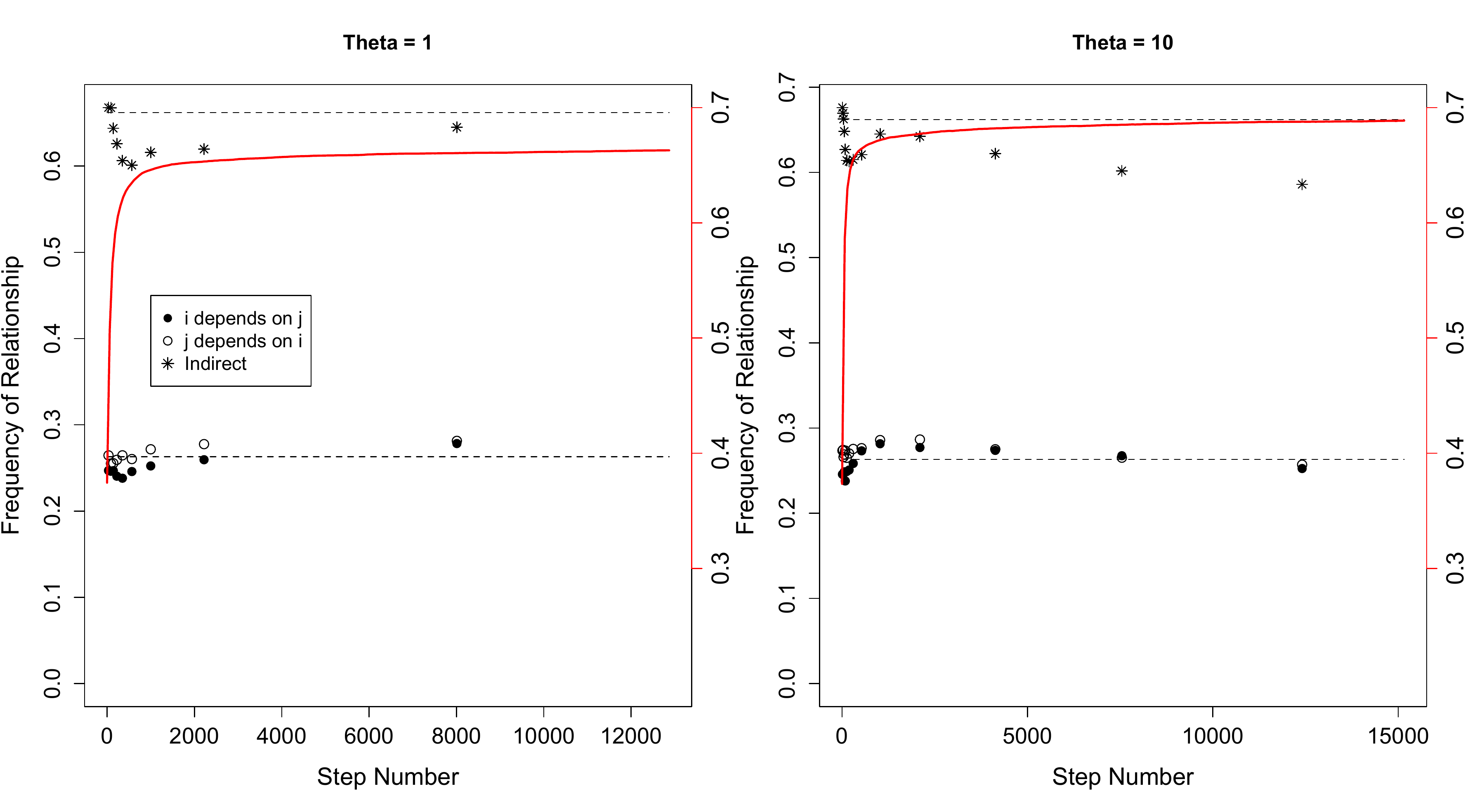} \caption{Observed frequency of directional epistatic interactions between substitution $i$ and its immediate successor $j$ for individual-based simulations with population sizes of 1000 and per-genome mutation rates of 0.001 and 0.01. A single pair may be counted as more than one type of epistasis, as in the case of reciprocal interactions. For $K > 1$, another class of epistasis is possible: both $i$ and $j$ may jointly influence a third site, which we refer to as an indirect interaction. The dashed lines depicts the predicted incidences if sites are chosen uniformly at random, as calculated in Eqs. 2-5. $N = 20$ and $K = 5$.} \end{figure}

\begin{figure}[!htb] \centering \includegraphics[width=14cm]{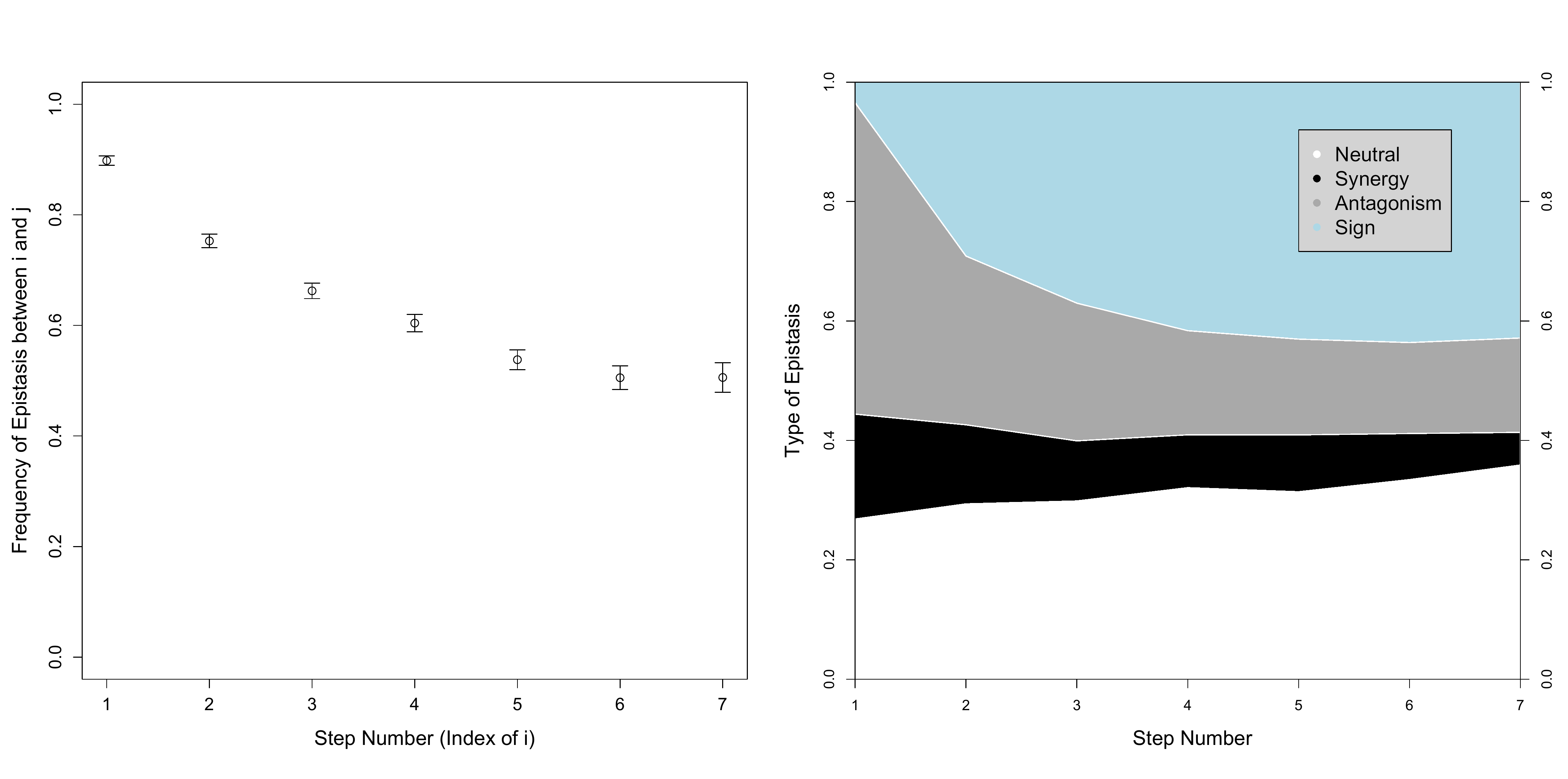}  \caption{Epistasis in individual-based evolution in a model of RNA folding. a) Observed frequency of epistatic interactions between substitution $i$ and its immediate successor $j$. Error bars indicate confidence intervals. b) Prevalence of four types of epistasis among those pairs of substitutions that interact. For both figures, population of 5000 individuals were simulated for 50,000 generations at a per-genome mutation rate of $2 \times 10^{-5}$. } \end{figure}

\begin{figure}[!htb] \centering \includegraphics[width=12cm]{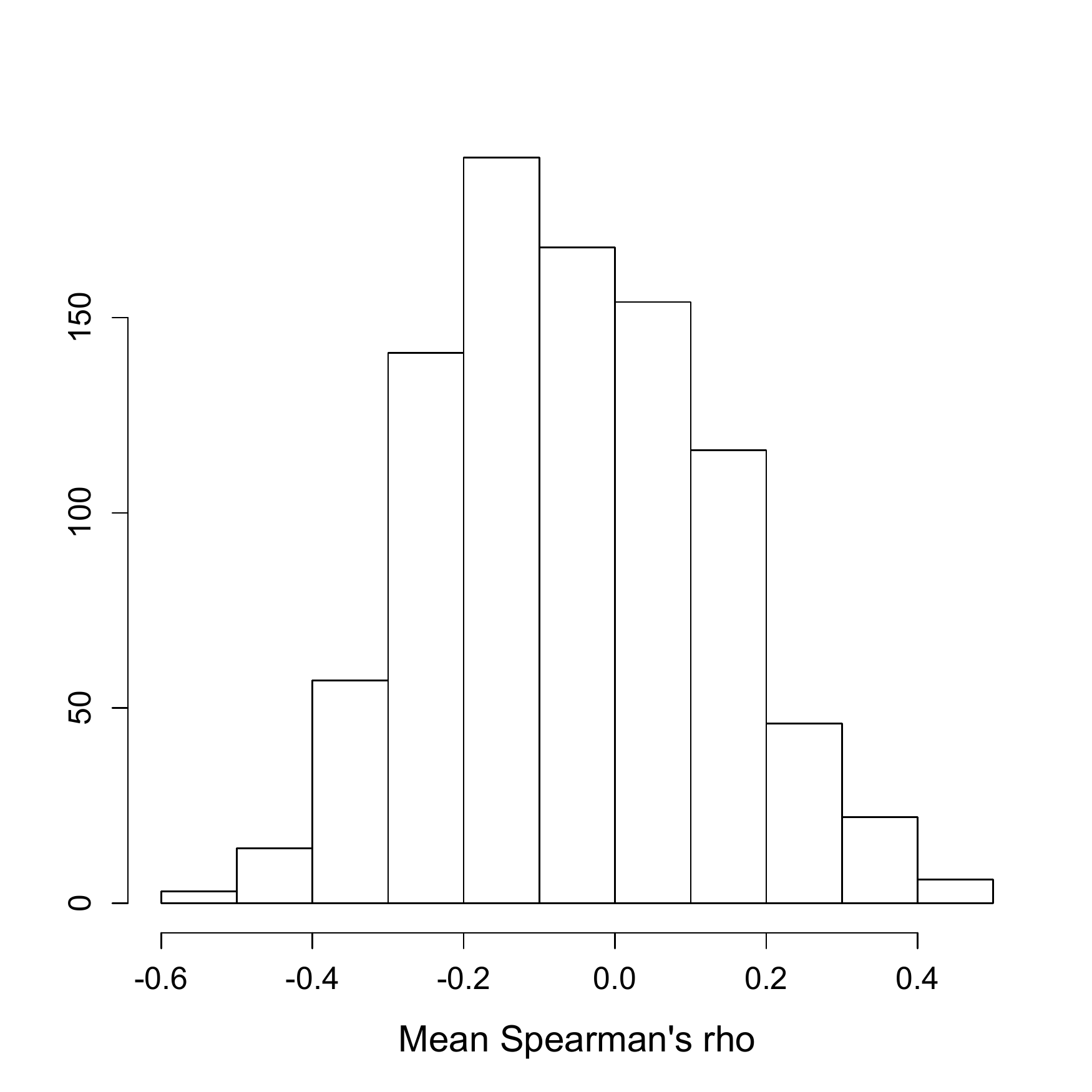} \caption{Spearman rank regression coefficients for the relationships between the fitness effect of a substitution and the fitness of the genetic background in which that substitution is made for adaptive walks which exceed multiplicative expectations at each step. Each substitution is tested against the sixteen genetic backgrounds comprising all combinations of the other four substitutions. Replicates are filtered to remove walks with too little epistasis among substitutions (see main text). $N = 20$, $K = 5$ and $A = 2$.}  \end{figure}

\begin{figure}[!htb] \centering \includegraphics[width=15cm]{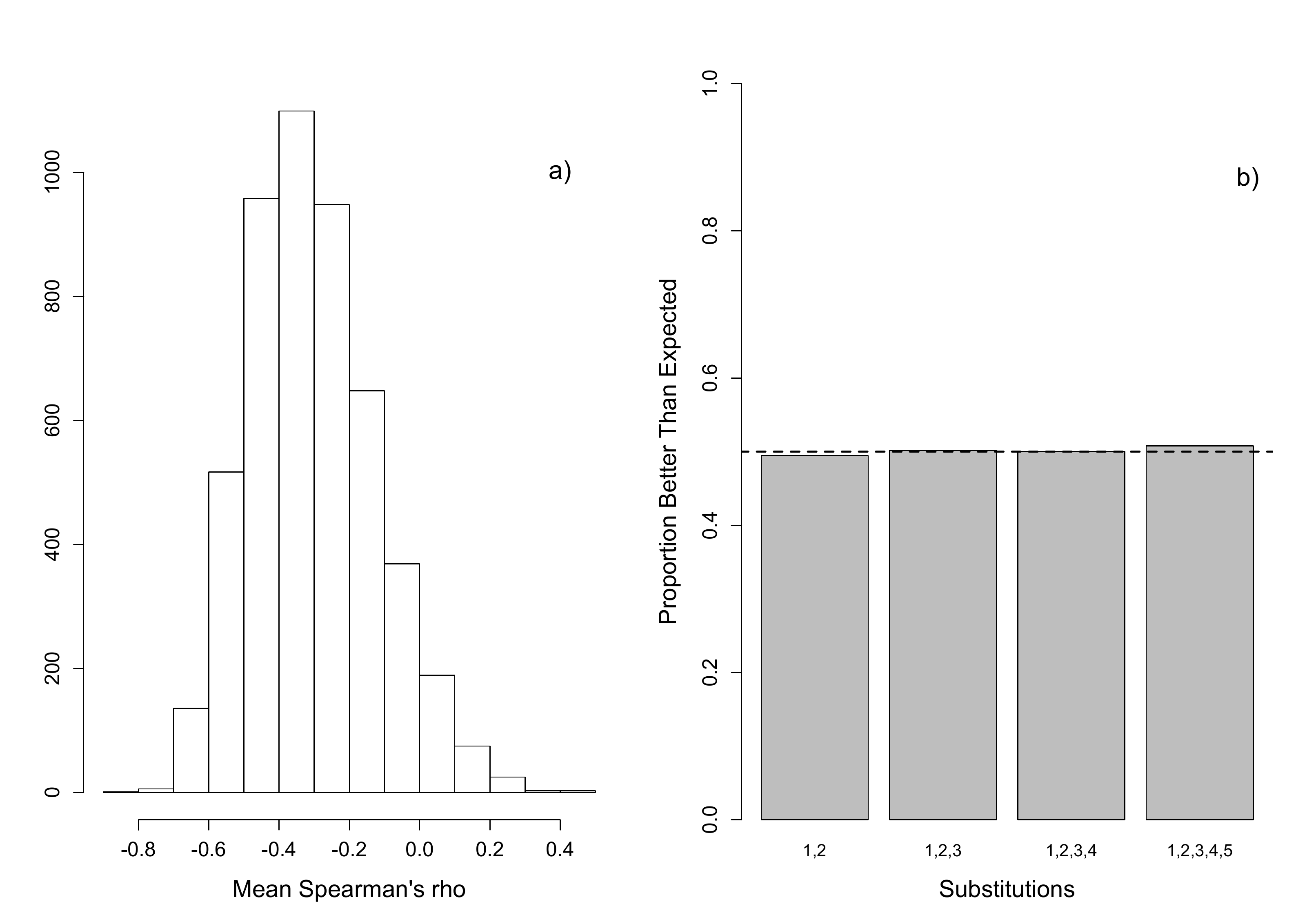} \caption{Two views of epistasis among the first five substitutions on random walks. a) Spearman rank regression coefficients for the relationships between the fitness effect of a substitution and the fitness of the genetic background in which that substitution is made. Each substitution is tested against the sixteen genetic backgrounds comprising all combinations of the other four substitutions. b) Proportion of combinations of substitutions along the line of descent which exceed the fitness expected from the fitness effects of the individual mutations in the ancestor. Expectations are computed according to Equations 7-10. Standard errors are less than 1\%. In both figures, replicates are filtered to remove walks with too little epistasis among substitutions (see main text) -- data show in panel (b) are additionally filtered to remove cases where $W_{12} = W_1 W_2/W$. $N = 20$, $K = 5$ and $A = 2$.} \end{figure}

\begin{figure}[!htb] \centering \includegraphics[width=15cm]{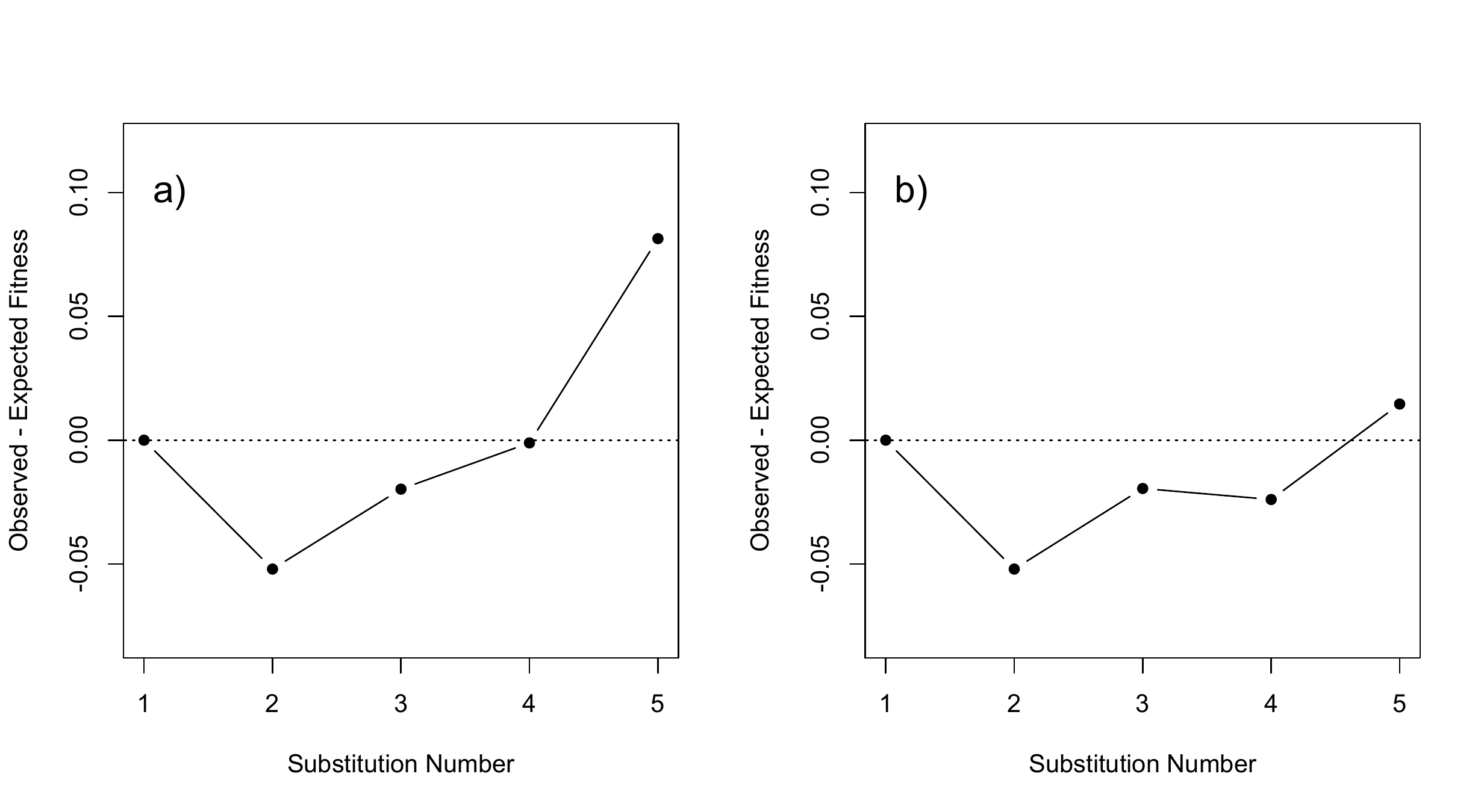} \caption{Epistasis deviation is initially negative, then becomes positive along the sequence of the first five substitutions observed in a microbial evolution experiment (Khan et al. 2011). The substituted loci are labeled $r$, $t$, $s$, $g$, and $p$ respectively. Let $w_r$ be the fitness of the mutant $r$ allele relative to the ancestor, $w_{rt}$ the fitness of the double mutant, and so on. Expected fitness can be calculated in at least two ways, each of which shows the same qualitative pattern. a) Expected fitness is computed as the product of the effects of each mutation in the ancestral genotype. For example, the deviation for substitution 3 is calculated as $\frac{w_{rts}}{w_{rt}} - w_s$. b) Expected fitness is computed with reference to the effect of the focal allele in the previous genetic background. For example, the deviation for substitution 3 is here calculated as $\frac{w_{rts}}{w_{rt}} - \frac{w_{rs}}{w_r}$.} \end{figure}

\end{document}